\documentclass[aps,twocolumn,pra,superscriptaddress]{revtex4-1}

\usepackage{amsmath,amsfonts,amssymb,graphicx,bbm,mathdots}
\usepackage[utf8]{inputenc}
\usepackage[usenames,dvipsnames]{xcolor}   % only for the comments

\usepackage{hyperref}

\usepackage{appendix}

\hypersetup{colorlinks=true,
						linkcolor=NavyBlue,        %color of internal links (change box color with linkbordercolor)
						citecolor=NavyBlue,        % color of links to bibliography
						filecolor=NavyBlue,      			 % color of file links
						urlcolor= Black,                  % color of external links}
}

\begin{document}

\newtheorem{example}{Example}

\newtheorem{remark}{Remark}
\newtheorem{problem}{Problem}
\newtheorem{theorem}{Theorem}
\newtheorem{corollary}{Corollary}
\newtheorem{definition}{Definition}
\newtheorem{proposition}{Proposition}
\newtheorem{lemma}{Lemma}
\newcommand{\proofend}{\hfill\fbox\\\medskip }
\newcommand{\proof}[1]{{\bf{Proof.}} #1 $\proofend$}
% use proof environment of amsthm instead...
\newcommand{\nn}{{\mathbbm{N}}}
\newcommand{\rr}{{\mathbbm{R}}}
\newcommand{\cc}{{\mathbbm{C}}}
\newcommand{\zz}{{\mathbbm{Z}}}
\newcommand{\mbp}{\ensuremath{\spadesuit}}
\newcommand{\je}{\ensuremath{\heartsuit}}
\newcommand{\jd}{\ensuremath{\clubsuit}}
\newcommand{\id}{{\mathbbm{1}}}
\newcommand{\me}{\mathrm{e}}
\newcommand{\mi}{\mathrm{i}}
\newcommand{\md}{\mathrm{d}}
\newcommand{\sg}{\text{sgn}}
\newcommand{\defeq}{\stackrel{\rm def}{=}}
\newcommand{\mc}[1]{\textcolor{blue}{[M.C.: #1]}}

\def\>{\rangle}
\def\<{\langle}
\def\({\left(}
\def\){\right)}

\newcommand{\ket}[1]{\left|#1\right>}
\newcommand{\bra}[1]{\left<#1\right|}
\newcommand{\braket}[2]{\<#1|#2\>}
\newcommand{\ketbra}[2]{\left|#1\right>\!\left<#2\right|}
\newcommand{\proj}[1]{|#1\>\!\<#1|}
\newcommand{\avg}[1]{\< #1 \>}

\newcommand{\einfuegen}[1]{\textcolor{PineGreen}{#1}}
\newcommand{\streichen}[1]{\textcolor{red}{\sout{#1}}}
\newcommand{\todo}[1]{\textcolor{blue}{(ToDo: #1)}}
\newcommand{\transpose}[1]{{#1}^t}

%\delimitershortfall=-2pt

\date{\today}
\title[]{Quantum Kibble-Zurek physics in long-range transverse-field Ising models}
\author{Ricardo Puebla}\email{r.puebla@qub.ac.uk}
\affiliation{Centre for Theoretical Atomic, Molecular, and Optical Physics, School of Mathematics and Physics, Queen's University, Belfast BT7 1NN, United Kingdom}
\affiliation{Institut f\"{u}r Theoretische Physik and IQST,
	Albert-Einstein Allee 11, Universit\"{a}t Ulm, 89069 Ulm, Germany}
\author{Oliver Marty}
\affiliation{Institut f\"{u}r Theoretische
	Physik and IQST, Albert-Einstein Allee 11, Universit\"{a}t Ulm,
	89069 Ulm, Germany}
\author{Martin B. Plenio}
\affiliation{Institut f\"{u}r Theoretische
	Physik and IQST, Albert-Einstein Allee 11, Universit\"{a}t Ulm,
	89069 Ulm, Germany}

\begin{abstract}
We analyze the quantum phase transitions taking place in a one-dimensional transverse field Ising model with long-range couplings that decay algebraically with distance. We are interested in the Kibble-Zurek universal scaling laws emerging in non-equilibrium dynamics and in the potential for the unambiguous observation of such behavior in a realistic experimental setup based on trapped ions. To this end, we determine the phase diagram of the model and the critical exponents characterizing its quantum phase transitions by means of density-matrix renormalization group calculations and finite-size scaling theory, which allows us to obtain good estimates for different range of ferro- and antiferromagnetic interactions. Beyond critical equilibrium properties, we tackle a non-equilibrium scenario in which quantum Kibble-Zurek scaling laws may be retrieved. Here it is found that the predicted non-equilibrium universal behavior, i.e. the scaling laws as a function of the quench rate and critical exponents, can be observed in systems comprising an experimentally feasible number of spins. Finally, a scheme is introduced to simulate the algebraically decaying couplings accurately by means of a digital quantum simulation with trapped ions. Our results suggest that quantum Kibble-Zurek physics can be explored and observed in state-of-the-art experiments with trapped ions realizing long-range Ising models.
\end{abstract}

\date{\today}

\maketitle

\section{Introduction}

The diversity of the macroscopic forms of matter has driven the interest in a comprehensive understanding of the underlying physical processes. At equilibrium, we distinguish different phases in which matter can organize itself, i.e., with different order, depending on the external conditions that predominate. Accordingly, each phase exhibits a specific qualitative behavior such as the solid, liquid and gas phase of matter, the paramagnetic and the ferromagnetic phases of a material or the degeneracy of ground states leading into spontaneous symmetry breaking, which can typically be quantified by an order parameter~\cite{Huang:87}. Similarly fascinating are the transitions from one phase to another when the external conditions change, which happen at certain values of external parameters such as the temperature or the strength of an applied magnetic field. The scientific advances taken in the past decades have led to an in-depth classification of phase transitions. An important class are continuous phase transitions which exhibit a continuous first derivative of the free energy and a discontinuity in the second (or higher order) derivative. Notably, for quantum phase transitions (QPTs), which take place at zero temperature, this condition is applied to the ground state energy~\cite{Sachdev:11}.

The Ising model plays an important role in understanding critical phenomena. Even though it was originally regarded as too simple to account for magnetism due to the absence of a classical phase transition in one dimension~\cite{Ising:25}, it is nowadays considered a cornerstone in classical and quantum statistical mechanics~\cite{Huang:87,Sachdev:11,Dutta:15}. Indeed, the transverse field Ising model (TIM) is arguably the simplest spin model which exhibits a continuous QPT, and constitutes the paradigmatic example in this realm~\cite{Sachdev:11,Dutta:15}.

An essential aspect of continuous phase transitions can be described within the powerful framework of universality: When approaching the critical point, the physical behavior of the system becomes universal, i.e., for example the typical length scale of correlations and the order parameter follow a power law determined by the critical exponents that characterize the phase transition. Such critical exponents, together with the dimensionality of the system define a universality class~\cite{Stanley:99}. In this manner, different systems that exhibit a phase transition belonging to the same universality class feature the same physical behavior close to the critical point. Furthermore, at a continuous QPT the energy gap often closes in a universal way~\cite{Sachdev:11}.

Interestingly, universal behavior at a continuous phase transition is also present in time-dependent processes far beyond the equilibrium condition. A framework which describes universality in a time-dependent scenario is the Kibble-Zurek mechanism (KZM) of defect formation~\cite{Kibble:76,Kibble:80,Zurek:85,Zurek:96}. A continuous phase transition typically occurs at a symmetry breaking phase transition. The subject of the KZM is the time evolution with the system initially in an equilibrium state of the symmetric phase, where, subsequently, by changing the external parameter in time across the critical point the system is forced to select a symmetry broken configuration. The KZM then predicts that, in the final state of the evolution, a variety of non-equilibrium quantities scale as a power of the rate at which the system traverses the critical point. Crucially, the exponents of the scaling relations are determined by the equilibrium critical exponents of the phase transition. Hence, the KZM reveals an intrinsic connection between the equilibrium properties and the dynamics of the system. While classical systems have originally been studied, it has later been shown that the KZM can also be applied to QPTs~\cite{Zurek:05,Damski:05,Dziarmaga:05,Polkovnikov:05,Kolodrubetz:12,Chandran:12,Polkovnikov:08} (see~\cite{delCampo:14} for a review).

In many cases, the scaling laws can be obtained by resorting to the adiabatic-impulse approximation. In this approximation, the state remains frozen when the external parameter is close to the critical point, i.e., it does not follow the evolution of the Hamiltonian. Outside of this stage of the evolution the system is changing adiabatically. Within this simplified picture, the transition between the two stages is supposed to be sharp at a certain value of the external parameter which sets the length scale of the correlations in the final state. The scaling laws predicted by the adiabatic-impulse approximation have been investigated theoretically in a number of classical settings~\cite{Laguna:97,Laguna:98,delCampo:10,Damski:10,Nigmatullin:16,Puebla:17b,Silvi:16} and found to be in good agreement with experimental results~\cite{Pyka:13,Ulm:13,Monaco:02,Navon:15,Beugnon:17}. The quantum Kibble-Zurek (QKZ) scaling laws, on the other hand, have been studied in~\cite{Zurek:05,Polkovnikov:05,Dziarmaga:05,Damski:05,Silvi:16}.

Similar to many other quantum many-body phenomena~\cite{Hohenberg:77,Schuetzhold:06,Polkovnikov:11,Eisert:15,Heyl:13,Heyl:18,Jurcevic:17,Zhang:17,Bernien:17}, the experimental progress of the last years opens up a new approach to study the QKZ mechanism (QKZM) in a controlled way by employing quantum simulators. However, the high degree of control and protection against noise that is required to experimentally realize the KZM in a quantum system represents a significant challenge. Experimental confirmations of the QKZM have only recently been achieved with Bose-Einstein condensates~\cite{Clark:16,Anquez:16}, using Rydberg atoms~\cite{Keesling:19}, or by directly simulating the Landau-Zener dynamics~\cite{Xu:14,Cui:16,Gong:16}. In addition, it is worth mentioning other aspects of universality arising in non-equilibrium dynamics, which have been recently identified in Refs.~\cite{Erne:18,Pruefer:18,Eigen:18}. Despite this progress, various aspects of non-equilibrium dynamics in isolated and open many-body quantum systems remain to be explored~\cite{Polkovnikov:11,Eisert:15}.

On the other hand, the transverse field Ising model with long range interactions has been realized with great success in various experiments with trapped ions, see, e.g.,~\cite{Kim:10,Smith:16,Islam:11,Friedenauer:08,Kim:09,Islam:13,Richerme:14,Cohen:15,Neyenhuis:17,Britton:12,SafaviNaini:18}, and has been the subject of recent theoretical studies of, both, ground-state properties~\cite{Koffel:12,Fey:16,Vodola:16,Sun:17,Defenu:17,Zhu:18} and non-equilibrium aspects such as entanglement dynamics~\cite{Hauke:13,Schachenmayer:13,Pappalardi:18}, the quantum Kibble-Zurek scaling~\cite{Jaschke:17} and other dynamical properties~\cite{Gong:13,Titum:18,Halimeh:17,Zauner:17,Homrighausen:17}. 

In this article, we analyze the scaling laws predicted by the QKZM in the one-dimensional long-range transverse field Ising model (LRTIM) with ferro- and antiferromagnetic couplings that decay algebraically with distance, and discuss the feasibility of observing the QKZM in a state-of-art trapped ion setup. To this end, first, the phase diagram of the QPT and the equilibrium critical exponents are calculated using density-matrix renormalization group (DMRG)~\cite{White:92,Schollwoeck:05,Schollwoeck:11}. Employing finite-size scaling theory~\cite{Fisher:72,Brankov:96}, we obtain good estimates of these quantities for different interaction ranges, which we corroborate by means of finite-size collapse. Further, we compare our results with previously published works, Refs.~\cite{Koffel:12,Fey:16,Vodola:16,Sun:17,Defenu:17,Zhu:18,Jaschke:17}. For the antiferromagnetic case, inconsistent results have been reported previously: Our findings support the hypothesis that the LRTIM remains within the universality class of the nearest-neighbor Ising model. On the other hand, for the ferromagnetic case, our results are in good agreement with previous studies. In the second part of the article, we numerically investigate the QKZM scaling laws for various quantities, such as the number of defects. Here, we employ the exact simulation of a system with up to 18 spins. For these system sizes, which are well within reach for current experiments with trapped ions, we find reasonably good agreement with the scaling predicted by the QKZM and the adiabatic-impulse approximation. Our results indicate therefore that QKZ physics can be readily accessed in state-of-the-art experiments. Finally, we discuss an efficient scheme to implement the time-evolution under the model with trapped ions using a digital quantum simulator that exploits the natural structure of the interactions that are available for this platform.

The article is organized as follows. In Sec.~\ref{sec:GSprop} we introduce the LRTIM and determine its phase diagram and equilibrium critical exponents. In Sec.~\ref{sec:QKZM} we first review the KZM of defect formation in the quantum regime, to then show that the predicted QKZ scaling laws can be observed in the LRTIM with experimentally feasible parameters. Finally, in Sec.~\ref{sec:expreal} we discuss the experimental realization of the LRTIM with trapped-ions using a digital quantum simulator approach, while the main conclusions are summarized in Sec.~\ref{sec:conclusions}.

\section{Equilibrium properties of the LRTIM}
\label{sec:GSprop}

We investigate the LRTIM of a one-dimensional system consisting of an even number $N$ of spin-1/2 particles. The Hamiltonian of the model can be written as (with $\hbar = 1$)

\begin{equation}
\label{eq:LRTIM_H}
\hat{H}(g) = \sum_{\substack{i,j = 1, \\ i < j}}^N J_{i,j} \hat{\sigma}_i^x \hat{\sigma}_j^x + g \sum_{i=1}^N \hat{\sigma}_i^z,
\end{equation}
with the strength of the transverse field $g \ge 0$, and algebraically decaying interactions
\begin{equation}
\label{eq:Jij_alpha}
J_{ij} = \frac{J_0}{|i-j|^{\alpha}},
\end{equation}
where $J_0 < 0$ ($J_0 > 0$) for ferromagnetic (antiferromagnetic) couplings and $\alpha \ge 0$. Here, $\hat{\sigma}_i^x,\hat{\sigma}_i^y,\hat{\sigma}_i^z$ denote the Pauli matrices acting on site $i$.  In the following we will choose $J_0 = \pm 1$. By controlling the parameter $\alpha$, different long-range interaction mechanisms can be described by the model, such as dipole-dipole ($\alpha = 3$) or van der Waals ($\alpha = 6$) interactions. Further, the limiting cases yield two important (solvable) models: First, in the limit $\alpha \rightarrow \infty$ the nearest-neighbor TIM is retrieved -- the paradigmatic example of a system undergoing a QPT~\cite{Sachdev:11,Dutta:15}. Second, for $\alpha = 0$ we have that $J_{ij} \equiv J_0$ and therefore the system is fully connected known as the Lipkin-Meshkov-Glick model~\cite{Lipkin:65}. In this case, the dimensionality of the system is $d=0$.

Independent of the exact form of the couplings, the model exhibits a $\mathbb{Z}_2$ symmetry associated with the invariance of the Hamiltonian Eq.~\eqref{eq:LRTIM_H} under spin-flip, i.e., since
$[\hat{\Pi},\hat{H}] = 0$, where
\begin{equation}
\hat{\Pi} = e^{i \frac{\pi}{2} \sum_{i=1}^N (\hat{\sigma}_i^z + \id)}
\end{equation}
denotes the parity operator. Thus, the Hilbert space is split into two subspaces with even and odd parity, respectively. Throughout the following, we will work within the positive parity subspace as it contains the ground state of $\hat{H}$ for sufficiently large $g$.

The ground state properties of the model depend crucially on the sign of $J_0$, which can be illustrated with the help of the nearest-neighbor TIM: For antiferromagnetic couplings and $g=0$ the ground state is a superposition of two staggered-magnetic ordered states $\ket{\rightarrow, \leftarrow, \rightarrow, ..., \rightarrow, \leftarrow}$ and $\ket{\leftarrow, \rightarrow, \leftarrow, ..., \leftarrow, \rightarrow}$, where $\hat{\sigma}^x \ket{\rightarrow} (\ket{\leftarrow}) = \ket{\rightarrow} (-\ket{\leftarrow})$  (recall that we consider $N$ even). In contrast, for ferromagnetic couplings the ground state is a superposition of the two fully-magnetic states $\ket{\rightarrow, ...,\rightarrow}$ and $\ket{\leftarrow, ...,\leftarrow}$. For $g \gg |J_0|$ the ground state becomes paramagnetic, i.e., a fully-polarized state $\ket{\downarrow, ...,\downarrow}$ where we introduced $\hat{\sigma}^z \ket{\uparrow}(\ket{\downarrow}) = \ket{\uparrow}(-\ket{\downarrow})$.

\begin{figure}[t]
	\includegraphics[angle=-90,width=0.499\textwidth]{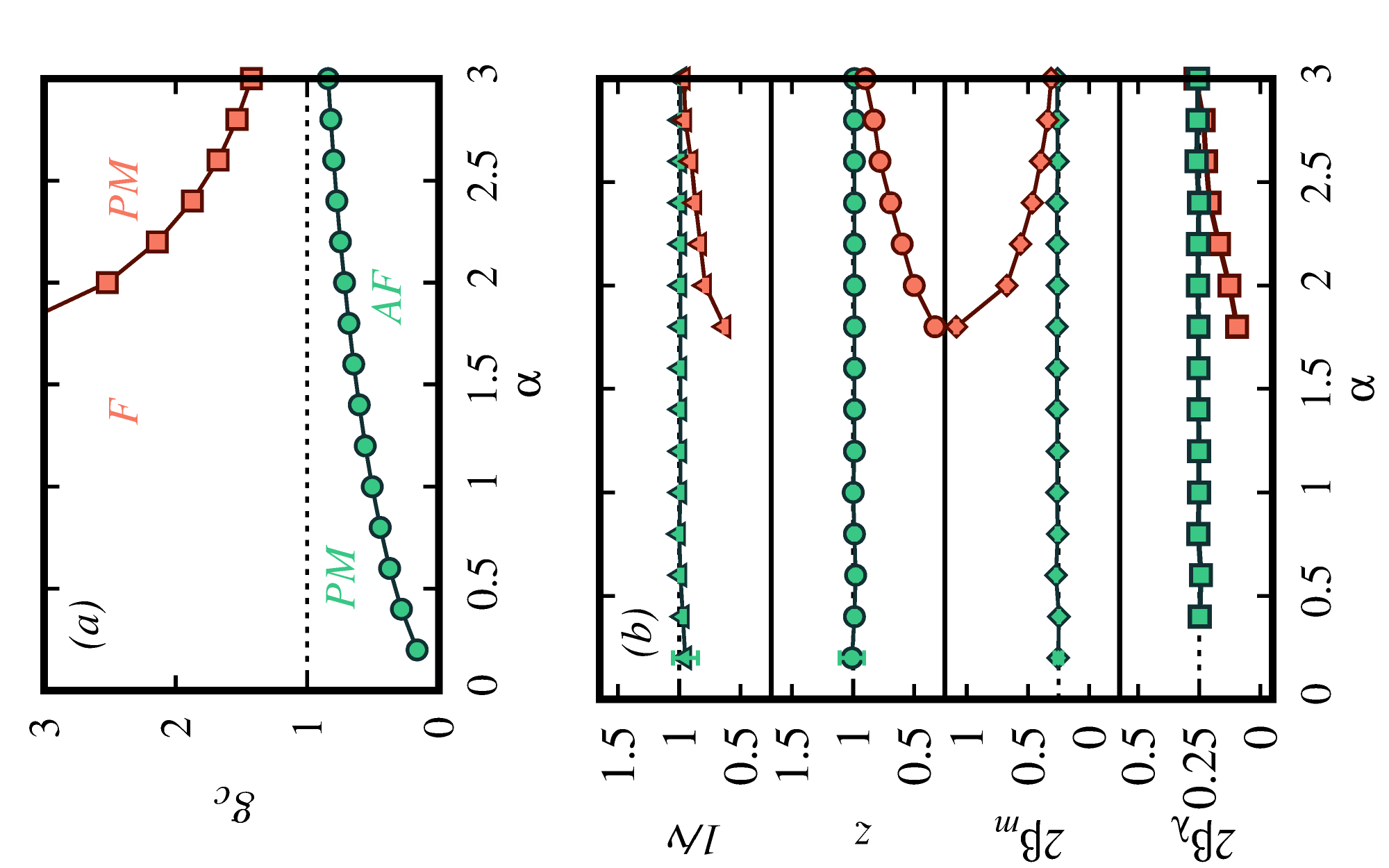}
	\caption{\label{fig:phasediag}(a) Phase diagram ($g$-$\alpha$) of the LRTIM: The circles (green) and squares (red) correspond to $g_c$ in units of $|J_0|$ for antiferromagnetic and ferromagnetic couplings, respectively, separating the paramagnetic phase (PM) from the antiferromagnetic (AF) and ferromagnetic (F). The dotted line shows the corresponding $g_c$ of nearest-neighbor case $\alpha \rightarrow \infty$. (b) From top to bottom, the critical exponents $1/\nu$, $z$, $2\beta_m$ and $2\beta_\lambda$. The dotted lines mark the values for $\alpha \rightarrow \infty$.}
\end{figure}

We now set out to determine the phase diagram of the ferro- and the antiferromagnetic LRTIM and characterize the universality classes for different values of the parameter $\alpha$. To this end, we employ that in the thermodynamic limit and close to the critical magnetic field strength $g_c$ various quantities are expected to show universal behavior. That is, for $g$ sufficiently close to $g_c$ the correlation length diverges as~\cite{Sachdev:11}
\begin{equation}
\label{eq:corrlength}
\xi \sim |g - g_c|^{-\nu},
\end{equation}
while the energy gap and the order parameter vanish as
\begin{equation}
\Delta \sim |g - g_c|^{z\nu},
\end{equation}
and (with $\zeta \in \{F,AF\}$)
\begin{equation}
\langle \hat{m}_\zeta \rangle \sim |g - g_c|^{\beta_m},
\end{equation}
respectively, where $\nu,z,\beta_m \ge 0$ are the critical exponents. Here the operator associated with the order parameter takes the form
\begin{equation}
\hat{m}_{F} = \frac{1}{N} \sum_{i=1}^N \hat{\sigma}_i^x,
\end{equation}
and 
\begin{equation}
\hat{m}_{AF} = \frac{1}{N} \sum_{i=1}^N (-1)^i \hat{\sigma}_i^x, 
\end{equation}
for ferro- and antiferromagnetic interactions, respectively.
Additionally, we determine the Schmidt gap $\Delta \lambda$~\cite{DeChiara:12,Lepori:13} as an indicator of the QPT, which is expected to behave as
\begin{equation}
\Delta \lambda \sim |g - g_c|^{\beta_{\lambda}},
\end{equation}
close to $g_c$.
The Schmidt gap is defined as the difference between largest two Schmidt coefficients for a specific bi-partition of the system. Here, we will consider a bipartition of the chain at the center into two blocks of equal size $N/2$. The Schmidt gap is related to the entanglement spectrum, that is, the eigenvalues of the reduced density matrix, which can exhibit critical behavior~\cite{Li:08} (however, it can be misleading, see~\cite{Chandran:14}). Note that the entanglement spectrum~\cite{Dalmonte:18}, Renyi entropies for pure states~\cite{Daley:12,Islam:15} and mixed state entanglement measures~\cite{Marty:16,Cramer:11,Marty:14} are experimentally accessible. 

For the nearest-neighbor interactions the critical point and the critical exponents can be determined exactly: One finds that with ferro- or antiferromagnetic interactions (these quantities coincide for both cases), the critical value is given by $g_c = 1$, while the exponents are $\nu = z = 1$ and $\beta_m = \beta_\lambda = 1/8$~\cite{DeChiara:12}. In particular, the order parameter and the Schmidt gap exhibit identical critical behavior.

To obtain the critical point and the exponents for the LRTIM we calculate $\xi$, $\langle \hat{m}_{\zeta}^2\rangle$ and $\Delta \lambda$ using DMRG. Recall that for any finite system with $N$ spins, the symmetry $\mathbb{Z}_2$ is not spontaneously broken and thus, constraining to the positive parity subspace leads into $\langle \hat{m}_{\zeta}\rangle=0$. We therefore calculate $\langle \hat{m}_{\zeta}^2\rangle$ instead which reveals critical behavior, and for which $\langle \hat{m}_{\zeta}^2\rangle\sim |g-g_c|^{2\beta_m}$ is expected. 
Here, to simulate the long-range interactions, we approximate the algebraic decaying couplings by a sum of exponentials~\cite{Murg:10}, see Appendix~\ref{app:DMRG} for further information. We consider system sizes of up to $N = 362$ spins with maximum bond dimension $200$, such that numerical convergence was attained (see Appendix~\ref{app:DMRG}).

In order to extrapolate the results obtained with systems of finitely many particles to the thermodynamic limit, we rely on finite size scaling theory~\cite{Fisher:72,Brankov:96}. Accordingly, any quantity $\mathcal{S}$ whose behavior in the thermodynamic limit $N \rightarrow \infty$ and sufficiently close to $g_c$ is of the form
\begin{equation}\label{eq:Sgamma}
\mathcal{S} \sim |g-g_c|^\gamma,
\end{equation}
(where $\gamma$ depends on $\mathcal{S}$)
can -- for any finite $N$ -- be written as
\begin{equation}
\label{eq:FSS}
\mathcal{S}(g,N) = N^{-\gamma/\nu} \phi_{\mathcal{S}}\!\left( (g-g_c) N^{1/\nu} \right) + o\left(N^{-\gamma/\nu}\right).
\end{equation}
Here, $\phi_{\mathcal{S}}$ is the finite-size scaling function associated with $\mathcal{S}$ which depends solely on the scaling variable $(g - g_c) N^{1/\nu}$. The term $o\left(N^{-\gamma/\nu}\right)$ represents sub-leading corrections to the scaling with $N$.

We can use the relation Eq.~\eqref{eq:FSS} to determine the phase diagram and the critical exponents. To find the critical field strength $g_c$, we rely on the Binder cumulant~\cite{Binder:81},
\begin{equation}
\label{eq:Binder}
B_\zeta = \frac{1}{2} \left( 3 - \frac{\langle \hat{m}^4_\zeta \rangle}{\langle \hat{m}^2_\zeta \rangle^2} \right)
\end{equation}
with $\zeta=F$ ($AF$) for ferromagnetic (antiferromagnetic) couplings. For the Binder cumulant, the relation Eq.~\eqref{eq:FSS} implies that close to $g_c$
\begin{equation}
\label{eq:Binderscal}
\begin{split}
B_\zeta(g,N) &= \frac{1}{2} \left( 3 - \frac{\phi_{\langle \hat{m}^4_\zeta \rangle} \left((g-g_c)N^{1/\nu}\right)}{\phi^2_{\langle \hat{m}^2_\zeta \rangle} \left((g-g_c)N^{1/\nu}\right)} \right) \\
&\equiv \phi_{B_\zeta}\left((g-g_c)N^{1/\nu}\right).
\end{split}
\end{equation}
That is, $B_\zeta$ becomes size-independent for $g = g_c$. Sub-leading corrections can result in small deviations from Eq.~\eqref{eq:Binderscal}. They are taken into account as follows: First, for two systems of sizes $N_1$ and $N_2$ we determine as a function of the product $N_1N_2$ the value $g^\ast(N_1 N_2)$ at which $B_\zeta(g^\ast,N_1) = B_\zeta(g^\ast,N_2)$. Then, using the hypothesis that $g^\ast(N_1 N_2) = g_c(1 + b (N_1 N_2)^{-\omega})$ for parameters $b$ and $\omega$, we fit $b, \omega$ and the critical point $g_c$ to the values of $g^\ast$ that we have obtained for several $N_1$ and $N_2$. See Appendix~\ref{app:GSprop} for further details.

We repeat these steps for different values of $\alpha$ and ferro- and antiferromagnetic couplings to obtain the phase diagram: Fig.~\ref{fig:phasediag}(a) shows $g_c$ as a function of $\alpha$ which sets the boundary between the paramagnetic (PM) and the ferromagnetic (F) or antiferromagnetic (AF) phase, respectively.

We remark that for ferromagnetic couplings we consider $\alpha \ge 1.8$ as for smaller values higher order finite-size corrections become more prominent and hinder the computation of the critical point $g_c$ with our approach. This is reflected by a decreasing value of $\omega$ as $\alpha \rightarrow 1$. In contrast, for antiferromagnetic couplings, similar behavior is observed for $\alpha \lesssim 0.4$, i.e., for much smaller values of $\alpha$, see also Appendix~\ref{app:GSprop}.

The critical exponents $z, \beta_m/\nu$ and $\beta_\lambda/\nu$ for ferro- and antiferromagnetic couplings are obtained from the scaling with $N$ of, respectively, the energy gap $\Delta$, the squared order parameter $\langle \hat{m}^2_{\zeta} \rangle$ (with $\zeta=F$ ($AF$) for ferromagnetic (antiferromagnetic) couplings) and the Schmidt gap $\Delta \lambda$ at the critical point $g_c$ using the relation Eq.~\eqref{eq:FSS}. On the other hand, to calculate the critical exponent $\nu$ we rely on the following scaling~\cite{Ferrenberg:91,Pelissetto:02,Angelini:14}
\begin{equation}
\label{eq:nuscal}
\frac{\left(\partial_g \langle \hat{m}^2_\zeta \rangle |_{g_c} \right)^2}{\partial_g \langle \hat{m}^4_\zeta \rangle |_{g_c}} \propto N^{1/\nu}.
\end{equation}
We refer to Appendix~\ref{app:GSprop} for the derivation of the previous expression, Eq.~\eqref{eq:nuscal}, as well as for an example showing the scaling (see Fig.~\ref{fig:app1}(d)).
In Fig.~\ref{fig:phasediag}(b) we show the results obtained numerically as a function of $\alpha$. We find that for ferromagnetic couplings all the critical exponents significantly change with $\alpha$. On the other hand, for antiferromagnetic couplings the critical exponents remain very close to the values of the nearest neighbor Ising universality class.

Further, we corroborate the validity of most of our results by means of finite-size collapse and verify the quality of each collapse employing a chi-squared test, see Appendix~\ref{app:FSC} for a detailed discussion. 

Notably, previous works have reported on the phase diagram of the LRTIM~\cite{Koffel:12,Sun:17,Zhu:18,Jaschke:17}, determined the product $z\nu$ based on linked-cluster expansion \cite{Fey:16} and calculated the critical exponents $\nu$ and $z$ via finite-size scaling~\cite{Sun:17,Zhu:18} and through renormalization group techniques~\cite{Defenu:17} (see also Ref.~\cite{Sperstad:12} for Monte-Carlo simulations of a spin chain interacting with a non-Ohmic bath and its connection with long-range models). The results on ferromagnetic interactions (including Refs.~\cite{Fey:16,Defenu:17,Zhu:18,Jaschke:17}) and the phase diagram of the antiferromagnetic interactions (determined in Refs.~\cite{Koffel:12,Vodola:16,Sun:17,Jaschke:17}) are in good agreement with our findings. On the other hand, the universality class of the QPT in the antiferromagnetic case is still under debate. In particular, the results based on DMRG reported in Ref.~\cite{Koffel:12} suggests that the model belongs to the nearest neighbor universality class for $\alpha \ge 9/4$ while for $\alpha < 9/4$ the critical exponents vary continuously with $\alpha$. Similarly, in Ref.~\cite{Fey:16} it is obtained that for $\alpha \ge 9/4$ the quantity $z\nu$ is close to 1 whereas it deviates from 1 for smaller $\alpha$ (although not consistently with Ref.~\cite{Koffel:12}). In contrast, the results presented in Ref.~\cite{Sun:17} strongly suggest that the nearest neighbor universality class holds for the antiferromagnetic model for any $\alpha > 0$, which agrees with our results (see~\cite{Fey:19} for a similar observation in two dimensions).

\begin{figure}[t]
	\includegraphics[width=0.49\textwidth]{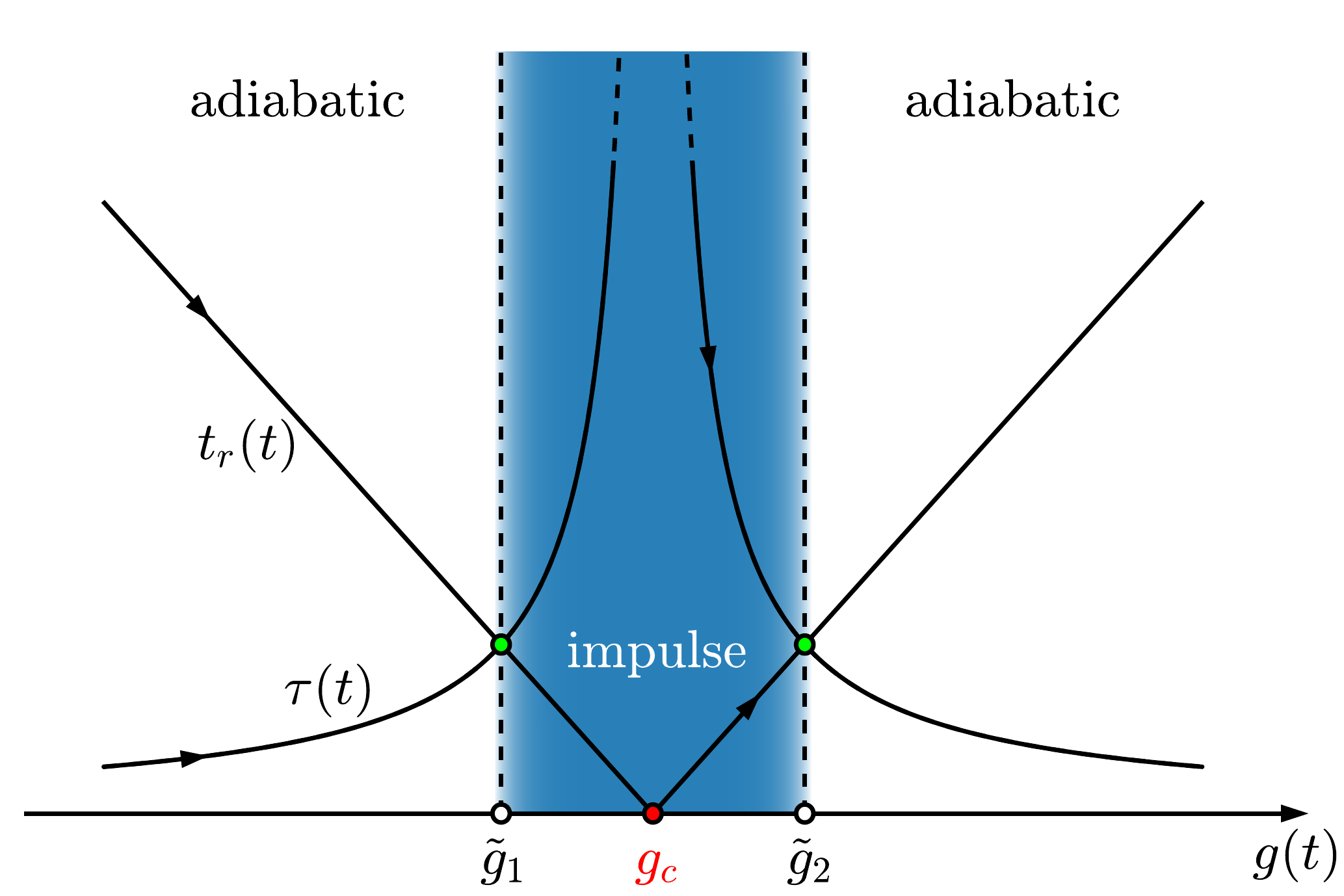}
	\caption{\label{fig:QKZscheme}Schematic representation of the adiabatic-impulse approximation that can be used to derive the Kibble-Zurek scaling. The points $\tilde g_1$ and $\tilde g_2$ mark the boundary of the impulse regime and are determined by the condition Eq.~\eqref{eq:AIboundary}.}
\end{figure}

\section{QKZM in the LRTIM}
\label{sec:QKZM}

In order to study the QKZM~\cite{Polkovnikov:05,Zurek:05,Dziarmaga:05,Damski:05} for the LRTIM, we consider the time evolution of the system where the external parameter is changed linearly in time according to
\begin{equation}
\label{eq:protocol}
g(t) = g_0 (1 - t/\tau_q),  \;\;\;\;\;\;\;\;\;\;   t \in [0,\tau_q],
\end{equation}
with $g_0 > g_c$ the value of the external parameter of the initial configuration and $\tau_q$ the duration of the evolution. In particular, throughout the ensuing, we consider the transition from the paramagnetic phase to the ferro- or antiferromagnetic phase. The state of the system that we are interested in is then given by the solution of the Schr\"odinger equation 
\begin{equation}
i\partial_t \ket{\psi(t)} = \hat{H}(g(t))\ket{\psi(t)},
\end{equation}
with the initial state $\ket{\psi(0)} = \ket{\phi_0(g_0)}$, where $\ket{\phi_0(g)}$ denotes the ground state of $\hat{H}(g)$.

Since the system undergoes a continuous phase transition it features a diverging relaxation time at the critical point $g_c$, that in the close vicinity of $g_c$ (and in the thermodynamic limit) reads
\begin{equation}
\tau(g) \simeq \tau_0 |g - g_c|^{-z\nu},
\end{equation}
where $\tau_0$ accounts for the microscopic details. As a consequence, the time evolution ceases to be adiabatic when it approaches the critical point and excitations in the instantaneous eigenbasis of the Hamiltonian $\hat{H}(g(t))$ are created. As predicted by the QKZM, the formation of these excitations follows scaling laws that are determined by the equilibrium critical exponents of the QPT and that are observable in the behavior of various quantities that depend on the time-evolved state.

%~\footnote{This excludes too fast quenches for which excitations are created even away from the QPT. Thus, QKZ scaling laws are expected to hold for $\tau_q\gtrsim 1/\omega_0$ where $\omega_0$ is a characteristic frequency of the model, while they will eventually break down when $\tau_q\gg \Delta^{-1}$ where $\Delta$ denotes the minimum energy gap.}. 

As mentioned in the Introduction, in order to derive the scaling laws, we rely on the adiabatic-impulse approximation which divides the dynamics into two regimes. In the adiabatic regime the state is supposed to adapt instantaneously to the Hamiltonian. On the other hand, in the impulse regime, the state does not follow the changes of $\hat{H}(g(t))$; it remains frozen. In the course of the time-evolution, adiabaticity is lost when the relaxation time $\tau(g)$ exceeds the timescale $t_r$ on which the external parameter is changing~\cite{Zurek:05,delCampo:14}. For the protocol Eq.~\eqref{eq:protocol}, the timescale $t_r$ takes the form
\begin{equation}
t_r(g) =  \frac{\tau_q}{g_0}|g-g_c|.
\end{equation}
The transition between the two regimes can thus be estimated to occur at values $\tilde g$ of the external parameter that satisfy the condition $\tau(\tilde g) = t_r(\tilde g)$. This yields the scaling
\begin{equation}
\label{eq:AIboundary}
|\tilde g - g_c| =  \left( \frac{\tau_0 g_0}{\tau_q} \right)^{1/(z\nu+1)} \sim \tau_q^{-1/(z\nu+1)},
\end{equation}
for the location and width of the impulse regime. We emphasize that this transition is just a working hypothesis, allowing for a heuristic derivation of the scaling with $\tau_q$. More rigorous scaling approaches can substantiate this derivation even in cases where the heuristic approach fails~\cite{Nikoghosyan:13}.
The QKZM now predicts that the average number of defects created during the time evolution is determined by the correlation length $\tilde{\xi} \equiv \xi(\tilde g)$ at the boundary, that is, 
\begin{equation}
\label{eq:ndef}
\left<\hat{n}_{\rm def}(\tau_q)\right> \sim \left(\frac{L}{\tilde\xi}\right)^d \sim \tau_q^{-d\nu/(z\nu+1)},
\end{equation}
where $L$ and $d$ denote the \emph{length} and dimensionality of the system, respectively, and where we used Eq.~\eqref{eq:AIboundary} and the relation Eq.~\eqref{eq:corrlength}. Here, the operator $\hat{n}_{\rm def}$ corresponds to the number excitations in the eigenbasis of the final Hamiltonian as noted above. This quantity may however be difficult to access experimentally for general long-range interactions. We therefore consider the average number of domains
\begin{equation}
\label{eq:ndomains}
\left<\hat{n}_{\rm do}\right> \equiv \frac{N+1}{2} \pm \sum_{i=1}^{N-1} \langle\hat{\sigma}_i^x \hat{\sigma}_{i+1}^x\rangle,
\end{equation}
which may be practically accessible. Here, the plus and the minus sign refer to ferro- and antiferromagnetic interactions, respectively. Since the operator $\hat{n}_{\rm do}$ defined in Eq.~\eqref{eq:ndomains} commutes with the Hamiltonian $\hat{H}(0)$ for any value of $\alpha$, we expect for $\left<\hat{n}_{\rm do}\right>$ the same scaling as for $\left<\hat{n}_{\rm def}\right>$ given in Eq.~\eqref{eq:ndef}~\cite{Polkovnikov:05,Deng:08}. Note that for the nearest-neighbor TIM with open boundary conditions as considered here, the two quantities are related as $\left<\hat{n}_{\rm def}\right> = \left<\hat{n}_{\rm do}\right> - 1$. In general, the final state is in a superposition of different spin configurations where each configuration comprises a particular number of domains, and thus, can be observed with a certain probability. We will come back to this in Sec.~\ref{ss:QKZsca}. In case of nearest-neighbor interactions, the domain sizes of the configurations that mainly contribute to the final state are limited by $\tilde \xi$, in agreement with the scaling Eq.~\eqref{eq:ndef}. However, this argument does in general not apply to long-range interactions. For example, for $\alpha = 0$, the Hamiltonian is invariant under permutations of the spins and therefore the number of the domains is not directly related to the number of defects.

\begin{figure}[t]
	\includegraphics[angle=-90,width=0.49\textwidth]{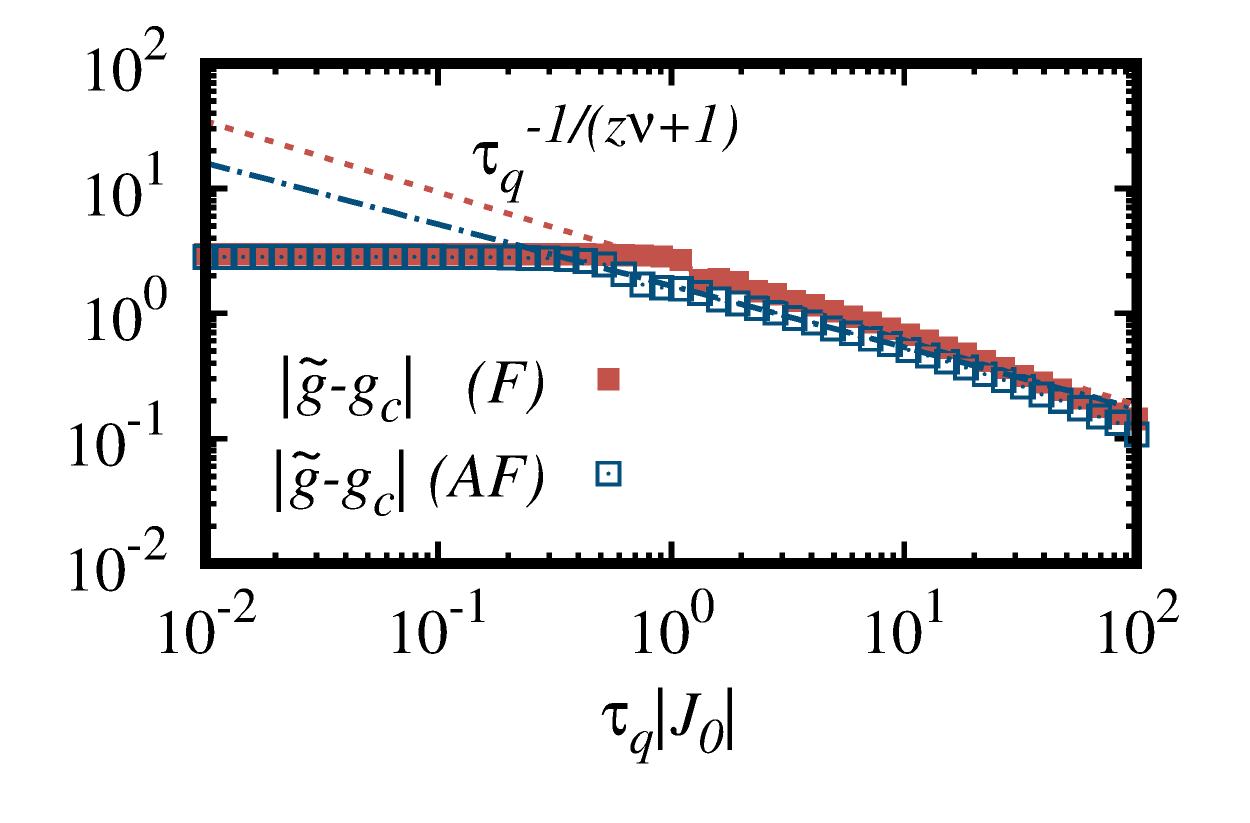}
	\caption{\label{fig:AI}Scaling of the size $|\tilde{g} - g_c|$ of the impulse regime in comparison with the prediction of the adiabatic-impulse approximation. The transition point $\tilde{g}$ is determined numerically as described in the main text. Shown are the results for ferro- (F) and antiferromagnetic (AF) couplings with $\alpha = 3$ and $N = 18$. For these two cases, the threshold is given by $\theta = 0.995$ and $0.999$, respectively. The lines correspond the fit $\propto \tau_q^{\mu}$ in the region $\tau_q \in [1,10]$, with $\mu = -0.49(1)$ and $-0.57(5)$, respectively.}
\end{figure}
\begin{figure*}[t]
	\includegraphics[angle=-90,width=0.999\textwidth]{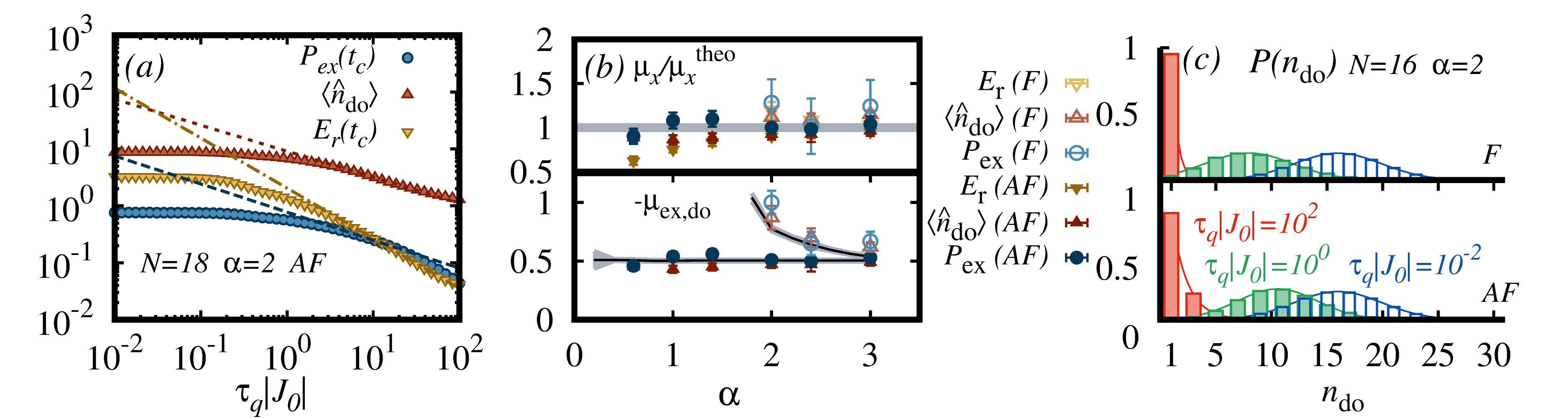}
	\caption{\label{fig:QKZscaling}(a) Scaling of excitation probability $P_{\rm ex}(t_c)$, the average number of domains $\left<\hat{n}_{\rm do}\right>$ and the residual energy $E_{\rm r}(t_c)$ for antiferromagnetic couplings with $\alpha = 2$ and $N = 18$ spins. The dashed, the dashed-dotted and the dotted lines show the corresponding fits to $\propto \tau_q^\mu$. Top panel in (b) shows the ratio between the fitted and the predicted theoretical values for the QKZ scaling exponents, $\mu_x/\mu_x^{\rm theo}$, for $x\in \{ {\rm do},{\rm ex},{\rm r} \}$, while the bottom panel shows the comparison of the fitted $\mu_{\rm ex,do}$ and the QKZ prediction with error bars shown as gray shaded area. (c) The probability distribution $P(n_{\rm do})$ of the number of domains $n_{\rm do}$ shows a Gaussian behavior ($\tau_q|J_0| = 1$ and $10^{-2}$) while it becomes Poissonian for adiabatic case where we observe $\langle \hat{n}_{\rm do}\rangle\approx 1$ (solid lines are best fits to $P(n_{\rm do})$) in agreement with~\cite{delCampo:18}.}
\end{figure*}

Further, we consider two additional quantities whose QKZ scaling can be predicted by similar arguments~\cite{Polkovnikov:05,deGrandi:10b,deGrandi:10,deGrandi:10c}: The excitation probability,
\begin{equation}
\label{eq:Pex}
P_{\rm ex}(t) = 1 - |\braket{\psi(t)}{\phi_0(g(t))}|^2
\end{equation}
and the residual energy,
\begin{equation}
\label{eq:Er}
E_{\rm r}(t) = \bra{\psi(t)}\hat{H}(g(t))\ket{\psi(t)} - E_0(g(t)),
\end{equation}
where $E_0(g) = \bra{\phi_0(g)}\hat{H}(g)\ket{\phi_0(g)}$. 
%When these quantities are measured at the end of the protocol where $g=0$, again the same scaling as for the number of defects is expected as these quantities (are the expectation values of operators that) commute with $H(0)$. 
%On the other hand, if $E_{\rm r}$ is measured at $t_c$ the scaling is of the form
The expected QKZ scaling of these quantities upon crossing the QPT are given by
\begin{align}
\label{eq:Pexscal}
P_{\rm ex}(\tau_q) \sim {\tau_q}^{-d\nu/(z\nu+1)},\\ \label{eq:Erscal}
E_{\rm r}(\tau_q) \sim {\tau_q}^{-d\nu/(z\nu+1)},
\end{align}
while right at the critical point, i.e. at $t_c=(1-g_0/g_c)\tau_q$ such that $g(t_c)=g_c$, their scaling becomes
\begin{align}\label{eq:Pexscalgc}
  P_{\rm ex}(t_c)&\sim \tau_q^{-d\nu/(z\nu+1)},\\
\label{eq:Erscalgc}
E_{\rm r}(t_c) &\sim {\tau_q}^{-(d+z)\nu/(z\nu+1)}.
\end{align}
We refer the reader to Refs.~\cite{deGrandi:10b,deGrandi:10,deGrandi:10c} for a derivation of such scaling laws based on the adiabatic perturbation theory and adiabatic-impulse approximation.   Notably, the residual energy Eq.~\eqref{eq:Er} belongs to a larger class of quantities of the form $\mathcal{S}_r(t) \equiv |\bra{\psi(t)}\hat{\mathcal{S}}\ket{\psi(t)} - \bra{\phi_0(g(t))}\hat{\mathcal{S}}\ket{\phi_0(g(t))}|$ that exhibit a QKZ scaling law, where $\hat{\mathcal{S}}$ denotes an operator for which there is critical behavior (i.e., $\mathcal{S}=\bra{\phi_0(g)}\hat{\mathcal{S}}\ket{\phi_0(g)}$ fulfills  Eq.~(\ref{eq:Sgamma})). Here, the scaling is given by $\mathcal{S}_r(\tau_q)\sim \tau_q^{-d\nu/(z\nu+1)}$ and $\mathcal{S}_r(t_c) \sim \tau_q^{-(d\nu+\gamma)/(z\nu+1)}$, where $\gamma$ is the equilibrium critical exponent of $\mathcal{S}$~\cite{Deng:08,deGrandi:10,deGrandi:10b,deGrandi:10c}. It is worth stressing that for $d=0$ systems QKZ scaling has been found only for ramps ending at the critical point~\cite{Hwang:15,Puebla:17,Defenu:18}.

Notably, in order to be able to observe the QKZ scaling laws, the duration $\tau_q$ (of the linear quench) is required to be in a certain range. If the duration of the quench is too short, the time-evolution is not adiabatic, even away from the critical point. On the other hand, for a finite system, $\ket{\psi(t)}$ remains in the ground state if $1/\tau_q$ is considerably smaller than the energy gap at the critical point. The time evolution is in this case is fully adiabatic to a good approximation, and thus the excitation probability and the residual energy are expected to scale as $\sim \tau_q^{-2}$~\cite{Polkovnikov:05,Rigolin:08,deGrandi:10b}. For more details on the influence that a finite system size has on the scaling, see Sec.~\ref{ss:noneqFSC}.

\begin{figure}[t]
	\includegraphics[angle=-90,width=0.49\textwidth]{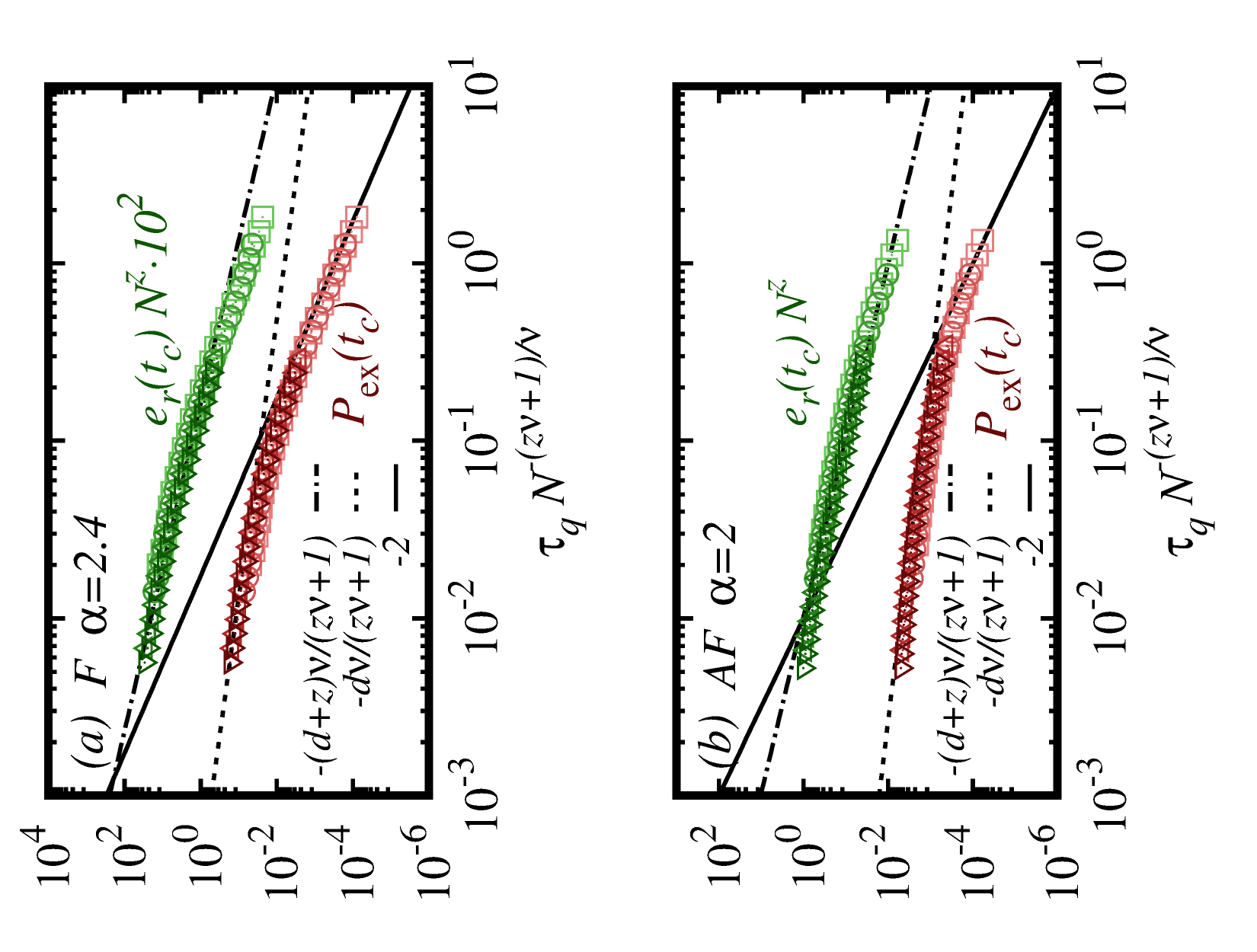}
	\caption{\label{fig:NoneqFSC} Non-equilibrium finite-size collapse for ferromagnetic (a) and antiferromagnetic couplings (b), with $\alpha = 2.4$ and $2$, respectively. The non-equilibrium scaling function $\bar{\phi}_{\mathcal{S}}(x=0,y)$ is plotted as a function of $y=N^{-(z\nu+1)/\nu}\tau_q$ for the residual energy $e_{\rm r}(t_c)\equiv E_{\rm r}(t_c)/N$ (multiplied by $10^{2}$ in (a) for a better illustration) and the excitation probability $P_{\rm ex}(t_c)$. We consider systems with $N = 8, 10, 16$ and $18$ (from light to dark color), and quench times $0.5 \le \tau_q|J_0| \le 10^2$. The lines show the expected adiabatic scaling $\tau_q^{-2}$ (solid) and the predictions by the QKZM, $\tau_q^{-(d+z)\nu/(z\nu+1)}$ (dash-dotted) and $\tau_q^{-d\nu/(z\nu+1)}$ (dotted) for $e_{\rm r}(t_c)$ and $P_{\rm ex}(t_c)$, respectively.}
\end{figure}

\subsection{Verification of the adiabatic-impulse approximation}

As discussed in the previous section, the Kibble-Zurek scalings are obtained from the adiabatic-impulse approximation which is a strong simplification. We therefore first investigate numerically the validity of this approximation for our model. To this end, in order to determine the transition point $\tilde g$, we measure the loss of adiabaticity of the time-evolution by the instantaneous ground-state fidelity
\begin{equation}
F(t) = |\braket{\psi(t)}{\phi_0(g(t))}|^2.
\end{equation}
When the system enters in the impulse regime, $F$ is expected to decrease suddenly. We thus identify the instant $t_{\theta}$ at which $F$ drops below a certain threshold $\theta$. We determine $t_{\theta}$ for different quench times $\tau_q$ to obtain $\tilde{g}(\tau_q) = g(t_{\theta}(\tau_q))$. 

For the LRTIM with $\alpha = 3$ and $N=18$ and, both, ferro- and antiferromagnetic couplings, the scaling of $|\tilde{g} - g_c|$ is shown in Fig.~\ref{fig:AI}. For this example, we employ the protocol Eq.~\eqref{eq:protocol} with $g_0 = 5$. In the figure, we further show an exponential fit $\propto \tau_q^\mu$ and the theoretically predicted scaling $\sim\tau_q^{\mu^{\rm theo}}$ with $\mu^{\rm theo} = -1/(z\nu+1)$ (with $z$ and $\nu$ from Sec.~\ref{sec:GSprop}). We find that the fitted exponent follows closely the theoretically predicted universal values: The ratio between the exponents employing the equilibrium and the dynamical approach is given by $\mu/\mu^{\rm theo} = 1.01(1)$ for ferromagnetic and  $\mu/\mu^{\rm theo} = 0.98(3)$ for antiferromagnetic couplings. 

For lower values of $\alpha$, however, finite-size effects have a more significant impact on the scaling laws. For example, for $\alpha = 0.6$ and antiferromagnetic couplings we find $\mu/\mu^{\rm theo} = 0.88(3)$, i.e., the values are no longer compatible within our error estimation. For ferromagnetic couplings the finite-size effects are even more prominent, as commented in Sec.~\ref{sec:GSprop}. In these latter cases, the numerically determined value of $\mu$ strongly depends on the selected threshold $\theta$.

\subsection{QKZ scaling laws}
\label{ss:QKZsca}
Next, we analyze the scaling laws of the excitation probability $P_{\rm ex}$ and the residual energy $E_{\rm r}$, defined in Eqs.~\eqref{eq:Pex} and~\eqref{eq:Er}, respectively. At the time $t_c$, these quantities are expected to follow the scaling given in Eqs.~\eqref{eq:Pexscalgc} and~\eqref{eq:Erscalgc}. Further we investigate the scaling of the number of domains after traversing the critical point, $\left<\hat{n}_{\rm do}\right> \sim \tau_q^{-d\nu/(1+\nu z)}$, where $d=1$ for $\alpha>0$. To compare the predicted exponents with the numerical data we again calculate the time evolution under the protocol Eq.~\eqref{eq:protocol}. %, where we determine $P_{\rm ex}$ and $E_{\rm r}$ at the instant $t_c$ and $n_{\rm do}$ at $\tau_q$. 
For different values of $\alpha$ and $N=18$, we consider quench times $\tau_q|J_0| \in [10^{-2},10^2]$ and determine the scaling law by fitting the numerical data to $\propto\tau_q^{\mu_x}$, with $x \in \{{\rm do},{\rm ex},{\rm r}\}$ labeling the three quantities, namely, $\left< \hat{n}_{\rm do}\right>$, $P_{\rm ex}(t_c)$ and $E_{\rm r}(t_c)$. In Fig.~\ref{fig:QKZscaling}(a) we show the typical behavior of these quantities as a function of $\tau_q$ for antiferromagnetic couplings with $\alpha = 2$. The exponential scaling is fitted in the range $5 \lesssim \tau_q|J_0| \lesssim 50$. The fitted exponents are found to be very close to the expected scaling, ruling out trivial adiabatic scaling. We can clearly distinguish the regime of sudden quenches ($\tau_q|J_0| \ll 1$) where the scaling deviates from the one predicted by the QKZM, and the quasi-adiabatic regime ($\tau_q|J_0| \gtrsim 1$). Not visible in the figure is the adiabatic regime which occurs for longer quench times, which is characterized by $\langle \hat{n}_{\rm do}\rangle \approx 1$ and the trivial scaling $\tau_q^{-2}$ of the $P_{\rm ex}$ and $E_{\rm r}$ as discussed in Sec.~\ref{sec:QKZM}.

 In Fig.~\ref{fig:QKZscaling}(b), top panel, we show the ratio between the fitted exponent and the QKZ scaling prediction, $\mu_x/\mu_x^{\rm theo}$ for $x \in \{{\rm do},{\rm ex},{\rm r}\}$, as a function of $\alpha$. The strongest deviation from the equilibrium exponent occurs for the residual energy for small $\alpha$. This suggests that finite-size effects are more pronounced in this quantity. In the lower panel of Fig.~\ref{fig:QKZscaling}(b) we plot the actual values for $\mu_{\rm do}$ and $\mu_{\rm ex}$ for ferro- and antiferromagnetic couplings. The solid dark lines and the gray shaded region correspond to $d\nu/(z\nu + 1)$ and its error for the equilibrium critical exponents $z$ and $\nu$ obtained in Sec.~\ref{sec:GSprop}.  Our results support previous studies of the QKZM for the antiferromagnetic case~\cite{Jaschke:17}. In contrast, for ferromagnetic couplings we find that the exponents depend strongly on $\alpha$, which is not in agreement with the results reported in~\cite{Jaschke:17}. See also~\cite{Dutta:17} for a non-trivial dependence of the QKZM scaling law on the range of the couplings for a different long-range interacting model. 

The presence of finite-size effects is furthermore visible in the form of the distribution of the number of defects or domains~\cite{delCampo:18}. That is, let $P(n_{\rm do})$ denote the probability of observing $n_{\rm do}$ domains at $\tau_q$. Then, it is expected that $P(n_{\rm do})$ forms a normal distribution if the quench times admit the Kibble-Zurek scaling, while it is of the form of a Poisson binomial distribution for slower quenches where the scaling breaks down due to finite-size effects. This behavior can precisely be observed in our model exemplified for $\alpha = 2$ and $N = 16$ in Fig.~\ref{fig:QKZscaling}(c): While for $\tau_q|J_0| = 10^{-2}$ and $1$ the normal distribution fits the numerical data very well, for $\tau_q|J_0| = 10^2$ the data displays an exponential decay $P(n_{\rm do})\propto e^{-n_{\rm do}}$. 

In addition, it is worth mentioning that the fitted exponents do not change significantly if the initial state is given by the fully-polarized state $\ket{\psi_0} = \ket{\downarrow,...,\downarrow}$ instead of the ground state $\ket{\phi_0(g_0)}$. This can be important to experimentally observe the scaling laws, since the fully-polarized state may be simpler to be prepared.

Finally, we note that quench times around $\tau_q |J_0|\lesssim 50$ have been achieved experimentally~\cite{Neyenhuis:17,Smith:16}. This, together with the above findings suggest that the scaling laws predicted by the QKZM can be tested in state-of-the-art experiments.

\subsection{Non-equilibrium finite-size collapse}
\label{ss:noneqFSC}

The quantum Kibble-Zurek predictions are based on singularities at the critical point that, strictly speaking, are only present in the thermodynamic limit. In any finite system, we rely on the fact that the quantities of interest approach the critical behavior in a sufficiently rapid and systematic manner. For equilibrium quantities a powerful tool to approach this problem is the finite-size scaling theory that we have employed in Sec.~\ref{sec:GSprop}. A similar strategy can be also applied to the non-equilibrium scenario~\cite{Kolodrubetz:12,Acevedo:14,Francuz:16,Chandran:12,Puebla:17b,Puebla:17}. 

According to the QKZM, the length scale relevant for the scaling of a quantity $\mathcal{S}$ with $\tau_q$ is given by the correlation length $\tilde{\xi}$ at $\tilde{g}$. For any finite system of length $L$, however, the correlation length cannot exceed the size of the system. Since 
\begin{equation}
\label{eq:QKZcoarse}
L/\tilde{\xi} \sim N|\tilde{g}-g_c|^\nu \sim N\tau_q^{-\nu/(z\nu+1)},
\end{equation}
we thus introduce a non-equilibrium finite-size scaling function $\bar{\phi}_\mathcal{S}(x,y)$ depending on two scaling variables~\cite{Francuz:16,Chandran:12,Puebla:17b,Puebla:17}  (cf. Eq.~\eqref{eq:FSS})
\begin{equation}\label{eq:Snoneq}
\mathcal{S}(t,N) = N^{-\gamma/\nu}\bar{\phi}_\mathcal{S}( (g(t)-g_c)N^{1/\nu},N^{-(z\nu+1)/\nu}\tau_q).
\end{equation}
Here, we suppose that the quantity $\mathcal{S}$ is intensive in order to obtain a size-independent scaling function $\bar{\phi}_\mathcal{S}(x,y)$ with $x=(g(t)-g_c)N^{1/\nu}$ and $y=N^{-(z\nu+1)/\nu}\tau_q$. Note that, for fully adiabatic dynamics, $\tau_q\rightarrow\infty$, the equilibrium scaling function $\phi_{\mathcal{S}}$ must be recovered, thus $\lim_{y\rightarrow \infty}\bar{\phi}_{\mathcal{S}}(x,y)=\phi_{\mathcal{S}}(x)$ (see for example~\cite{Francuz:16,Puebla:17}). The scaling function $\bar{\phi}_\mathcal{S}(x,y)$ is therefore a non-equilibrium generalization of $\phi_{\mathcal{S}}(x)$, and it is expected to be valid when the loss of adiabaticity occurs close to the QPT~\cite{Francuz:16,Puebla:17,Acevedo:14}. That is, Eq.~\eqref{eq:Snoneq} does not hold for sudden or too fast ramps $\tau_q|J_0|\lesssim 1$. On the other hand, for a fixed value of $x_f$, the size-independent function $\bar{\phi}_\mathcal{S}(x_f,y)$ contains the QKZ scaling ($y\ll 1$) and the trivial quadratic scaling ($y\gg 1$). Following a similar argument as in Eq.~\eqref{eq:QKZcoarse}, resorting to the adiabatic-impulse approximation, the relaxation time close to the QPT scales as $\tau\sim \tau_q^{z\nu/(z\nu+1)}$, while its maximum value for a finite system follows $\tau\sim N^{z}$. Hence, when both are comparable, that is, when $y=\tau_q N^{-(z\nu+1)/\nu}\approx O(1)$, one expects a crossover between QKZ and the trivial $\tau_q^{-2}$ scaling regimes.

In order to illustrate this, we compute the non-equilibrium quantities after a time $t_c$ under the protocol Eq.~\eqref{eq:protocol} such that $g(t_c)=g_c$ and thus, $x_f=0$. In this case, it follows $\mathcal{S}(t_c,N)N^{\gamma/\nu}=\bar{\phi}_\mathcal{S}(0,N^{-(z\nu+1)/\nu}\tau_q)$. We show the finite-size collapse for $e_{\rm r}(t_c)\equiv E_{\rm r}(t_c)/N$ and $P_{\rm ex}(t_c)$ for two representative cases, namely $\alpha = 2.4$ for ferromagnetic and $\alpha = 2$ for the antiferromagnetic couplings, plotted in Fig.~\ref{fig:NoneqFSC}(a) and (b), respectively. It is worth noting that the good collapse of the data further supports that universality already plays a significant role in the dynamics of systems comprising a few number of spins.

\section{Experimental realization using a digital quantum simulator}
\label{sec:expreal}

Ising couplings with tunable interactions can experimentally be implemented with trapped ions. The theoretical proposal~\cite{Porras:04,Deng:05} was followed by experiments comprising two~\cite{Friedenauer:08} and, more recently, several tens of individually controllable ions realizing complex many-body physics, see, e.g., Refs.~\cite{Smith:16,Richerme:14,Neyenhuis:17,Lanyon:17,Friis:18}. In these setups, the energy levels of each ion that encode the qubit states are connected via either an optical \cite{Jurcevic:14,Jurcevic:17,Lanyon:17,Friis:18} or a microwave transition~\cite{Smith:16,Zhang:17,Zhang:17b,Neyenhuis:17,Olmschenk:07}. A state dependent force is obtained by applying a bichromatic laser field with frequencies of opposite detunings $\pm \mu$ from the qubit transition frequency. The couplings of the resulting phonon-mediated effective spin-spin interaction are of the form
\begin{equation}
\label{eq:Jijions}
J_{ij}^{\rm e} =  \Omega_i \Omega_j\frac{k_0^2}{2M}\sum_{m=1}^N \frac{b_{i,m}b_{j,m}}{\mu^2 - \omega_m^2},
\end{equation}
where $M$ is the mass of the ions, $b_{i,m}$ are the amplitudes of the $m$-th normal mode and $\omega_m$ the associated normal-mode frequency~\cite{James:98}. The Rabi frequency of the $i$-th ion  $\Omega_i$ and the wave number $k$ are parameters of the laser field whose exact form depends on the setup that is used. Additionally, a transverse magnetic field can be generated via an asymmetric detuning of the laser frequencies (AC-Stark shift) or with an additional laser field  in resonance with the two qubit states.

The range of the couplings $J_{ij}^{\rm e}$ can be adjusted by changing the laser detuning $\mu$ from infinite range when $\mu$ is tuned close to the center-of-mass mode frequency $\omega_{\rm com}$ to dipole-dipole interactions when $\mu$ is sufficiently far from any phonon mode frequency. The detuning can also be set to a value between these two cases such that the couplings approximately follow an algebraically decaying function as in Eq.~\eqref{eq:Jij_alpha}; but the deviation
from an ideal algebraic decay is typically strong and the quality of the approximation is not controllable. Therefore in order to study long-range interactions in a systematic way with trapped ions, it may be important to simulate the time evolution of the Ising model with couplings $J_{ij}$ given in Eq.~\eqref{eq:Jij_alpha} with higher precision than it is possible by relying only on the similarity of these couplings to the naturally occurring interactions $J_{ij}^{\rm e}$. This can be achieved using a digital quantum simulation~\cite{Lloyd:96}. So far, experimental implementations of digital quantum simulations with trapped ions employed infinite range interactions to couple the spin degrees of freedom~\cite{Lanyon:11,Lanyon:13}. Although it is possible to couple only specific pairs of spins by canceling some of these interactions using single-spin laser pulses~\cite{Lanyon:13}, such an approach is generally inefficient. In contrast, in the following we introduce a scheme that can be used to implement any type of Ising couplings with trapped ions by means of a sequence of at most $\mathcal{O}(N)$ gates. In our scheme, the interactions between the spins are created by exploiting the form of the couplings $J_{ij}^{\rm e}$ instead of using individual two-qubit gates. As an example, we focus here on the implementation of algebraically decaying couplings $J_{ij}$ and leave the detailed investigation of other coupling forms and of the efficiency of the scheme compared to other implementations for future work.

In order to describe our approach more specifically, let us consider the time evolved state $\ket{\psi(t)}$ of the model with a time-dependent magnetic field strength  $g(t)$ as it can be used for adiabatic ground-state preparation or to verify the QKZM. Notably, for the following discussion, $g(t)$ is not required to be linear as in Eq.~\eqref{eq:protocol}. To simulate the time evolution with a digital quantum simulator we divide it into a sequence of unitary operations using the Magnus method~\cite{Blanes:09} as
\begin{equation}
\label{eq:Magnusexp}
\ket{\psi(t)} \approx \hat{U}(t_{n-1},t_n) \cdots \hat{U}(t_0,t_1) \ket{\psi_0},
\end{equation}
with $t_{l+1} = t_{l} + \Delta t$, $t = t_n = n\Delta t$ and $\ket{\psi_0} \equiv \ket{\psi(0)}$ denoting
the initial state. Here, on every interval $[t_l,t_{l+1}]$, the time evolution is approximated by means of a time-independent Hamiltonian,
\begin{equation}
\hat{U}(t_l,t_{l+1}) = e^{-i \Delta t \hat{H}(g(t_l+\Delta t/2))}.
\end{equation}
This can further be expanded by the Lie-Trotter product formula: Suppose that we have a decomposition of the Hamiltonian as $\hat{H}(g) = \sum_{k=1}^q \hat{H}_k(g)$ for some operators $\hat{H}_k$, $k = 1,\ldots,q$. With such a decomposition, the time evolution operator $\hat U$ can be written as
\begin{equation}
\label{eq:LieTrotter}
\hat{U}(t_l,t_{l+1}) \approx \hat{U}_1^{(l)} \cdots \hat{U}_q^{(l)},
\end{equation}
where $\hat{U}_k^{(l)} = \exp[-i\Delta t \hat{H}_k(g(t_l+\Delta t/2))]$. The errors introduced in the two approximations above, Eqs.~(\ref{eq:Magnusexp}) and~(\ref{eq:LieTrotter}), can exactly be bounded and are in both cases of order $(\Delta t)^2$. Note that other methods to implement the above discretization steps can be used to improve the overall error of the time evolution~\cite{Wiebe:11}.
\begin{figure}[t]
    \includegraphics[width=0.49\textwidth]{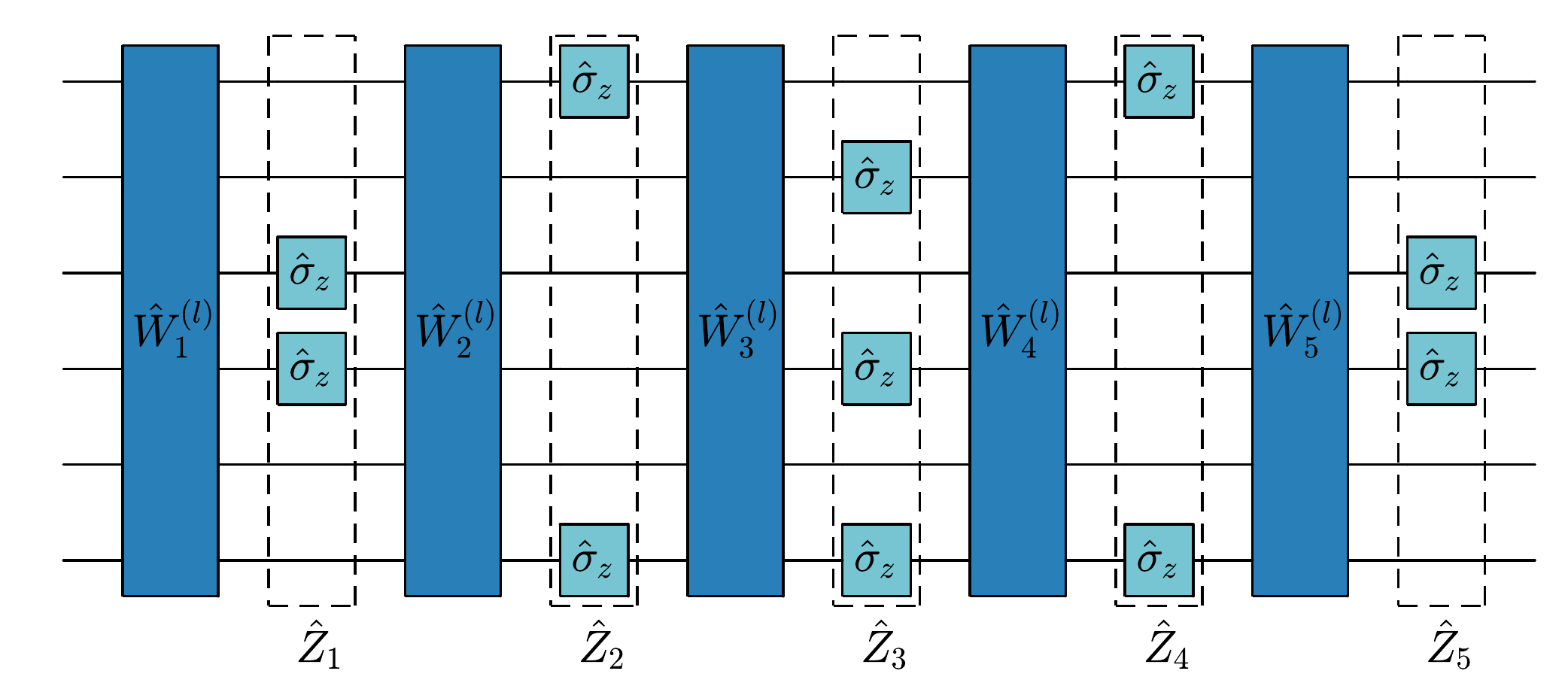}
    \caption{\label{fig:quantumcircuit}A sequence of $\mathcal{O}(N)$ gates that can be implemented using trapped ions to simulate the evolution under the TIM with algebraically decaying couplings as in Eq.~\eqref{eq:Jij_alpha}. The quantum circuit which is depicted describes an approximation of the operator $\hat{U}(t_l,t_{l+1})$, see Eq.~\eqref{eq:stepcircuit} and the main text for the definition of the unitaries $\hat{W}_k^{(l)}$. In this example $N=6$ and $\alpha = 1$.}
\end{figure}

We now set out to find an explicit form of the expansion Eq.~\eqref{eq:LieTrotter} that can be implemented with trapped ions. For the Hamiltonian $\hat H$ which is given in Eq.~\eqref{eq:LRTIM_H}, at first sight this expansion may only seem possible with a quantum circuit that employs two-body gates. However, we show now that it can be realized using a quantum circuit of depth at most $2(N-1)$ which consists of a combination of single-site operations and a global time evolution under the naturally occuring interactions Eq.~\eqref{eq:Jijions}. Our scheme is thus an example of a hybrid digital-analog simulation of a quantum system~\cite{Lamata:17,Lamata:18,Arrazola:16}. To derive the desired gate operations explicitly, we define the matrix $K$ whose entries are given by 
\begin{equation}
K_{ij} = \begin{cases} w & i = j, \\ J_{ij} & i\ne j. \end{cases}
\end{equation}
We can choose $w$ in order that $K$ is positive semi definite with smallest eigenvalue 0. The eigendecomposition of $K$ then takes the form $K = \sum_{k=1}^{N-1} \Lambda_k \vec{v}_k\vec{v}_k^{\rm T}$ with $\Lambda_k \ge 0$ the eigenvalues of $K$ that we label in decreasing order. Clearly, the Ising model with couplings given by $K_{ij}$ and $J_{ij}$ are equivalent since the diagonal elements of $K$ have no influence on any physical properties of the model. With this, a decomposition of the Hamiltonian as in Eq.~\eqref{eq:LieTrotter} is obtained if we define
\begin{equation}
\hat{H}_{k}(g) = \sum_{\substack{i,j = 1, \\ i < j}}^N \Lambda_k (\vec{v}_k)_{i}(\vec{v}_k)_{j} \hat{\sigma}_i^x \hat{\sigma}_j^x + \frac{g}{N-1} \sum_{i=1}^N\hat{\sigma}_i^z,
\end{equation}
for $k = 1,\ldots,N-1$.
Crucial for a decomposition into gates that can \textit{experimentally} be realized with trapped ions is the fact that the free evolution under $\hat{H}_k$ can be implemented by combining the evolution under the Ising model with couplings $J_{ij}^{\rm e}$ with local spin flips. Indeed, this can be achieved by the following three steps: First we apply spin-flips to all sites with $(\vec{v}_k)_i < 0$ by means of the unitary operator
\begin{equation}
\hat{\tilde{Z}}_{k} = \bigotimes_{\substack{i=1, \\ (\vec{v}_k)_i < 0}}^N \hat{\sigma}_i^z.
\end{equation}
Notably, since all entries in $K$ are strictly positive, we can assume that $(\vec{v}_1)_i \ge 0$ and hence $\hat{\tilde{Z}}_1 = \id$.
In the second step, we employ the evolution under the Ising model which depends on the detuning $\mu$ and the Rabi frequencies $\Omega_i^k$ where the latters are different for the $N-1$ factors in the Trotter decomposition. For this, we choose $\mu = \omega_{\rm com} + \delta$ with $\delta \ll |\mu - \omega_m|$ for all normal mode frequencies $\omega_m \ne \omega_{\rm com}$, and 
\begin{equation}
\Omega_i^k  = \Omega_0\frac{\left(2M|\Lambda_k|\right)^{1/2}}{k_0} (\vec{v}_k)_i,
\end{equation}
for some factor $\Omega_0$. Here, we remark that in order to avoid phonon excitations we require for all normal mode frequencies that $|\mu - \omega_m| \gg \eta_{i,m} \Omega_i^k$, with $\eta_{i,m} = b_{i,m} k_0(2M\omega_m)^{-1/2}$ the Lamb-Dicke parameter. This condition may be fulfilled by adjusting the laser intensity accordingly. Additionally, a transverse field term may be added by one of the standard methods for trapped ions mentioned above. We then define the time evolution operator $\hat{W}_k^{(l)} = \exp[-i (\Delta t/\Omega_0) \hat{H}_k^{\rm e}(g(t_l+\Delta t/2))]$ with
\begin{equation}
\hat{H}_{k}^{\rm e}(g) = \sum_{\substack{i,j = 1, \\ i < j}}^N J_{ij}^{\rm e}[k] \hat{\sigma}_i^x \hat{\sigma}_j^x + \frac{\Omega_0 g}{N-1} \sum_{i=1}^N\hat{\sigma}_i^z,
\end{equation}
for $k = 1,\ldots,N-1$, where $J_{ij}^{\rm e}[k]$ denotes the couplings $J_{ij}^{\rm e}$ given in Eq.~\eqref{eq:Jijions} with Rabi frequencies $\Omega_i = \Omega_i^k$. Finally, to complete the $k$-th trotterization step the operator $\hat{\tilde{Z}}_k$ is applied again, i.e., we have that $U_k^{(l)} \approx \hat{\tilde{Z}}_k W_k^{(l)} \hat{\tilde{Z}}_k$. The overall protocol to approximate the time evolution operator $\hat{U}$ using trapped ions thus reads
\begin{equation}
\label{eq:stepcircuit}
\hat{U}(t_l,t_{l+1}) \approx \hat{W}_1^{(l)} \hat{Z}_1  \cdots \hat{W}_{N-1}^{(l)} \hat{Z}_{N-1},
\end{equation}
where $\hat{Z}_{k} = \hat{\tilde{Z}}_{k}\hat{\tilde{Z}}_{k+1}$ for $k = 1,\ldots,N-2$ and $\hat{Z}_{N-1} = \hat{\tilde{Z}}_{N-1}$. Note that the gates $\hat{Z}_k$ consist only of single-site operations that can be applied simultaneously, while $\hat{U}_k$ addresses the ions globally. As an example, the quantum circuit to simulate the evolution under algebraically decaying couplings with $\alpha=1$ is depicted in Fig.~\ref{fig:quantumcircuit} for $N=6$ ions.

\section{Conclusions}
\label{sec:conclusions}

In summary, we have analyzed the QPTs taking place in the one-dimensional Ising model with algebraically decaying long-range interactions. We have determined the phase diagram for ferromagnetic and antiferromagnetic couplings as a function of the parameter $\alpha$, which accounts for the algebraically decaying couplings. For different values of $\alpha$, we have computed the equilibrium critical exponents of the QPTs in order to characterize their universality class. These results are obtained performing detailed simulations, based on DMRG calculations and finite-size scaling theory. We found that, while the paramagnetic-ferromagnetic QPT changes universality class for $1.8 \le \alpha \le 3$, the QPT for the frustrated case remains in its short-range universality class for $0.4 \le \alpha \le 3$. Our results partially support the findings of previous works. In addition, we have determined the critical exponents for the order parameter and Schmidt gap, which obey the same trend as $z$ and $\nu$. The correctness of the critical point and exponents are tested by means of finite-size collapse. Having obtained the phase diagram and critical exponents, we tackled the non-equilibrium dynamics resulting from traversing the QPT by linearly changing the magnetic field strength in time. We focused on the emerging scaling laws as a function of the quench rate, as predicted by the KZM of defect formation in a quantum system. To this end, we have first verified the adiabatic-impulse approximation and then analyzed the scaling in the average number of domains, the residual energy and the excitation probability for various $\alpha$. As we show, the fitted exponents are very close to the predicted ones. We again corroborate our findings by means of finite-size scaling for the non-equilibrium states.
Finally, we have discussed the experimental realization with trapped ions and proposed a scalable scheme to efficiently simulate the dynamics using a digital quantum simulator. 

Our findings show that the KZM of the LRTIM can be observed in system comprising only a few tens of spins. In view of the current technological advances, the paradigmatic QKZM represents a phenomenon that can be useful to benchmark quantum simulators as well as to unveil novel physical properties of many-body systems. In particular, our results suggest that QKZM can be explored in state-of-the-art trapped-ion settings which provide a fully controllable platform that naturally exhibits long-range interactions, whose properties may be hard to simulate on a classical device already for moderate system sizes.

\acknowledgements
R. P. thanks G. De Chiara for enlightening discussions and acknowledges the support by the SFI-DfE Investigator Programme (grant 15/IA/2864). We acknowledge helpful correspondence with B.~P. Lanyon. This work was supported by the ERC Synergy grant BioQ and the authors acknowledge support by the state of Baden-W\"urttemberg through bwHPC and the German Research Foundation (DFG) through grant no INST 40/467-1 FUGG.

\appendix

\section{Details of the DMRG calculations}
\label{app:DMRG}

 \begin{figure*}[t]
	\includegraphics[angle=-90,width=1\textwidth]{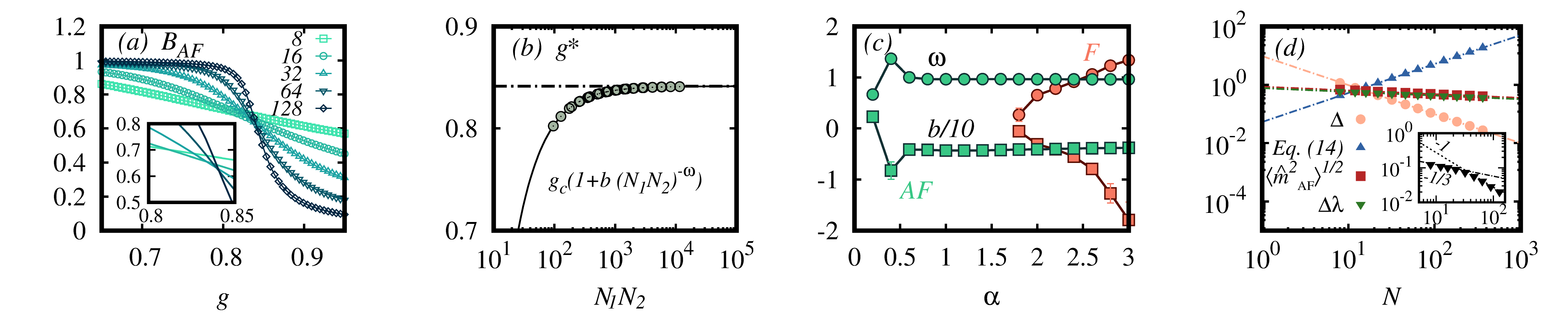}
	\caption{\label{fig:app1}(a) The Binder cumulant for different numbers of spins $N$ as a function of $g$. In the example shown in the figure, the couplings are antiferromagnetic with $\alpha = 3$. According to Eq.~\eqref{eq:Binderscal} in the main text, the curves intersect at the critical point $g_c$. Sub-leading order corrections result in deviations from this behavior, see inset. (b) The critical value for the example of (a) is obtained by fitting the values $g^\ast(N_1N_2)$ to the function Eq.~\eqref{eq:gc_crit}. The value $g^\ast(N_1N_2)$ is determined when two Binder cumulants, with system sizes $N_1$ and $N_2$, intersect. (c) The fitted values for the parameters $b$ (squares) and $\omega$ (circles) for the ferro- (red) and the antiferromagnetic (green) model as a function of $\alpha$. (d) The critical exponents $z$, $\nu$, $\beta_{m}$ and $\beta_\lambda$ result from the scaling of the energy gap $\Delta$, Eq.~\eqref{eq:nuscal}, $\langle \hat{m}_{\zeta}^2 \rangle^{1/2}$ (for $\zeta \in \{F,AF\}$) and the Schmidt gap $\Delta\lambda$. In the figure we show again the results for antiferromagnetic couplings with $\alpha = 3$. The inset shows another example of the scaling of the gap $\Delta \sim N^{-z}$ for antiferromagnetic couplings with $\alpha=0.2$. As guides to the eye, the lines represent the scalings $\propto N^{-1/3}$ (dashed) and $\propto N^{-1}$ (dotted).}
\end{figure*}

  \begin{figure*}[t!]
	\includegraphics[angle=-90,width=0.969\textwidth]{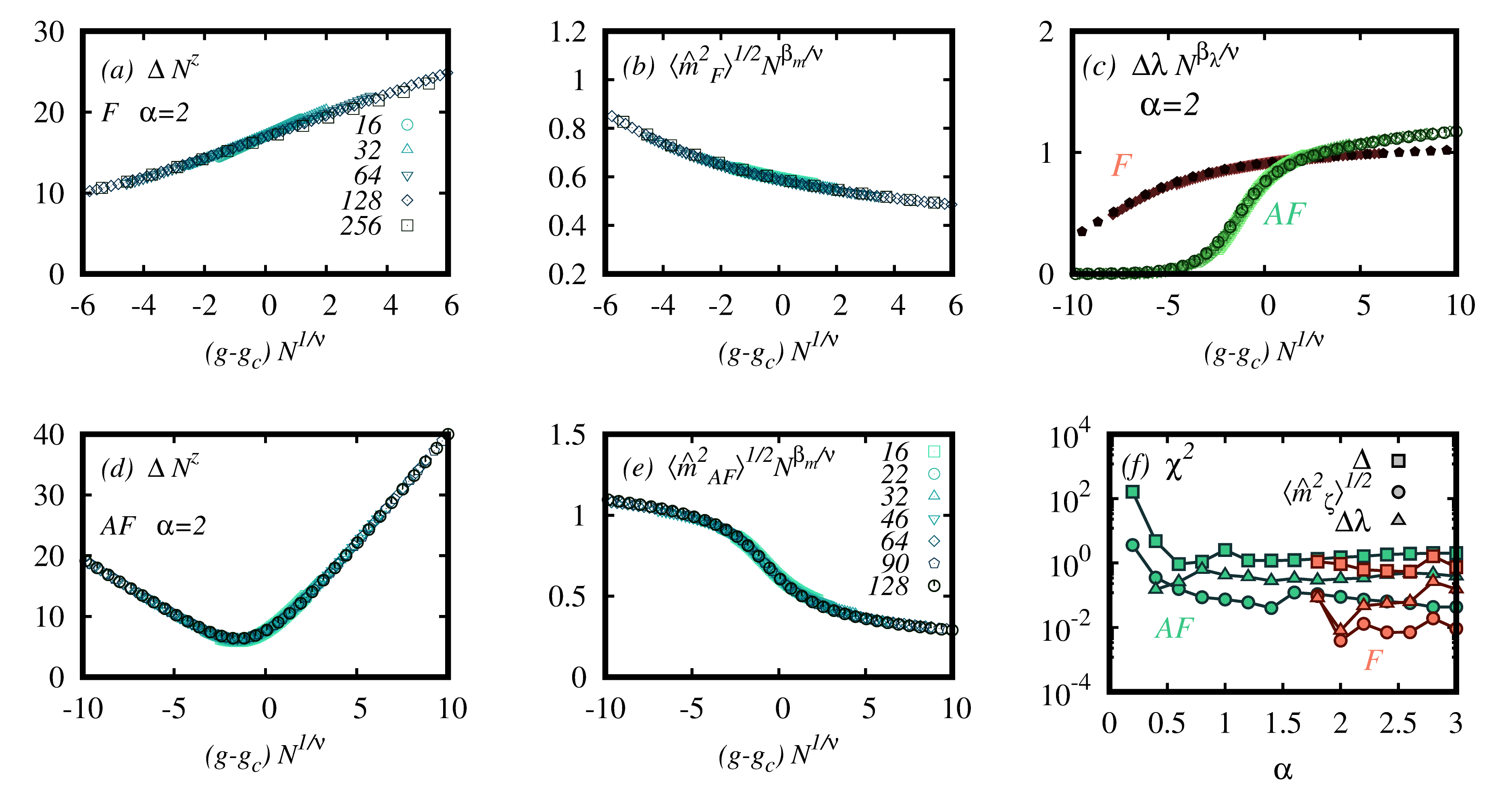}
	\caption{\label{fig:app2}(a)-(e) Finite-size collapse for the energy gap $\Delta N^z$, the order parameter $\langle \hat{m}_\zeta^2\rangle^{1/2} N^{\beta_m/\nu}$ and the Schmidt gap $\Delta\lambda N^{\beta_\lambda/\nu}$ as a function of $(g-g_c)N^{1/\nu}$ for ferro- and antiferromagnetic couplings with $\alpha = 2$. (f) The values $\chi^2$ of the chi-squared test which quantify the goodness of the collapse $\Delta$ (squares), $\langle \hat{m}_\zeta^2 \rangle^{1/2}$ (circles), for $\zeta \in \{F,AF\}$, and $\Delta\lambda$ (triangles).}
\end{figure*}
In order to find the ground state we employ a standard single-site variational matrix-product state method~\cite{Schollwoeck:05,Schollwoeck:11}. For this, the Hamiltonian has to be
expressed as a matrix-product operator (MPO) with a bond dimension that is sufficiently small to store it. One approach to obtain a suitable MPO representation is to expand the couplings in terms of a sum of exponentials~\cite{Murg:10} which we review in the following. That is, formally, one seeks for parameters $\{c_i\}_{i=1}^n$ and $\{\lambda_i\}_{i=1}^n$ with
\begin{equation}
\frac{1}{k^\alpha} \approx \sum_{i=1}^n c_i \lambda_i^k.
\end{equation}
In order to find these parameters, we can first consider the economy-sized QR-decomposition $A = VR$, where $F$ is an $(N-n+1) \times n$ matrix whose entries are given by $A_{i,j} = 1/(i+j-1)^\alpha$. The parameters $\{\lambda_i\}_{i=1}^n$ are choosen to be the eigenvalues of $V_1^+ V_2$, where $V_1$ ($V_2$) is the matrix containing the first (last) $(N-n)$ rows of $V$. Here, $V^+$ denotes the pseudo-inverse of a matrix $V$. The coefficients $\{x_i\}_{i=1}^n$ can then be found by a standard least-square optimization. In the numerical simulations, we utilize $n = 10$ terms which results in a approximation of the couplings of the order $\mathcal{O}(10^{-6})$ or better. 

Also, it can readily be shown~\cite{Schollwoeck:11} that an operator of the form 
\begin{equation}
\label{eq:exp_op}
\sum_{\substack{i,j=1, \\ i<j}}^N \lambda^{i-j} \hat{\sigma}_i^x \hat{\sigma}_j^x + g \sum_{i=1}^N \hat{\sigma}_z
\end{equation}
can be expressed as an MPO with bond dimension 3. Further, the sum of $n$ operators Eq.~\eqref{eq:exp_op} exhibits and MPO representation of bond dimension at most $3n$.

The convergence of the variational algorithm is verified by monitoring the truncation error and the smallest value of the entanglement spectrum  upon a bi-partition of size $N/2$, denoted $\lambda_m$. In particular, we choose the bond dimension such that $\lambda_m\leq 10^{-6}$ for $\alpha\geq 0.8$. For $\alpha<0.8$, our maximum bond dimension of $200$ provides $\lambda_m\leq 10^{-5}$.  

%The convergence of variational algorithm we verify by monitoring the truncation error and the variance $(\Delta H)^2$ which is a standard procedure for these methods. For the values of $\alpha$ that we study in this work we observe the errors are of order $10^{-6}$. 

\section{Additional information to the derivation of the ground state properties}
\label{app:GSprop}

It follows a description of the method employed to locate the critical point $g_c$ and determine the critical exponents. As commented in the main text, we resort to the Binder cumulant $B_\zeta$ with $\zeta\in\left\{F, AF \right\}$, Eq.~\eqref{eq:Binder}, to locate the critical point $g_c$ for various $\alpha$ values. For two different system sizes $N_1$ and $N_2$, the Binder cumulants intersect at $g^\ast$, thus providing an estimate of the actual $g_c$. An example is shown in Fig.~\ref{fig:app1}(a). As $N_1 N_2 \rightarrow \infty$, the intersection will occur closer to $g_c$, and hence, by fitting these estimates to~\cite{Angelini:14}
\begin{equation}
\label{eq:gc_crit}
g^\ast(N_1N_2) = g_c(1+b(N_1N_2)^{-\omega})
\end{equation}
we obtain precise results for $g_c$, see Fig.~\ref{fig:app1}(b). It is worth stressing that the exponent $\omega$ gives account for higher-order corrections (sub-leading finite-size scaling). In Fig.~\ref{fig:app1}(c) we show the fitted parameters, which indicate that the ferromagnetic LRTIM exhibits stronger sub-leading corrections than its antiferromagnetic counterpart for $\alpha \lesssim 2.4$. The fits for $\alpha \lesssim 0.4$ however become unstable, indicating that larger system sizes are required to obtain reliable estimates. At the critical point $g_c$, we can obtain the critical exponents through finite size scaling. From Eq.~\eqref{eq:FSS}, up to sub-leading corrections, we expect the scaling $\mathcal{S}(N,g_c) \propto N^{-\gamma/\nu}$ at the QPT for a quantity $\mathcal{S}$ as described in the main text. The critical exponent $\nu$ however must be determined in a different manner. For $\zeta\in\left\{F,AF\right\}$, we achieve this by computing the derivative of $\langle \hat{m}^{2n}_\zeta \rangle$,
\begin{equation}
\frac{\partial\langle \hat{m}^{2n}_\zeta \rangle}{\partial g} = N^{-2n \beta_m/\nu} \frac{\partial \phi_{\langle\hat{m}^{2n}_\zeta\rangle}}{\partial g}\!\left((g-g_c)N^{1/\nu} \right).
\end{equation}
Here, have used the finite-size scaling function, i.e., 
\begin{equation}
\langle \hat{m}^{2n}_\zeta\rangle = N^{-2n\beta_m/\nu} \phi_{\langle\hat{m}^{2n}_\zeta\rangle}((g-g_c)N^{1/\nu}),
\end{equation}
as we are interested in the case $g \approx g_c$. At the critical point, it follows
\begin{equation}
\begin{split}
\log \frac{\partial \langle \hat{m}^{2n}_\zeta \rangle}{\partial g}\bigg|_{g_c} &= \frac{1-2n\beta_m}{\nu} \log N + \log \frac{\partial \phi_{\langle \hat{m}^{2n}_\zeta \rangle}(x)}{\partial x}\bigg|_{x = 0} \\
&= \frac{1 - 2n\beta_m}{\nu} \log N + {\rm const.}
\end{split}
\end{equation}
Hence, $\nu$ can be obtained from the following expression:
\begin{equation}
2 \log \frac{\partial \langle \hat{m}^2_\zeta \rangle}{\partial g}\bigg|_{g_c} - \log \frac{\partial \langle \hat{m}^4_\zeta \rangle}{\partial g}\bigg|_{g_c} = \frac{1}{\nu}N + {\rm const.},
\end{equation}
 which is equivalent to Eq.~\eqref{eq:nuscal}. An example of the scaling of the different quantities is shown in Fig.~\ref{fig:app1}(d). In the inset of Fig.~\ref{fig:app1}(d), the energy gap $\Delta(N,g_c)$ is displayed for $\alpha = 0.2$ and antiferromagnetic interactions to illustrate the increasing difficulty to determine the exponent $z$ for small values of $\alpha$. Note that $\Delta$ seems to scale as $N^z$ with $z \approx 1/3$ (mean-field value) up to $N \approx 46$, shifting to $z \approx 1$ with increasing $N$.
 
 \section{Finite-size collapse}
 \label{app:FSC}
We verify the validity of the numerically calculated critical exponents via the finite-size collapse of $\Delta$, $\langle \hat{m}^2_\zeta \rangle^{1/2}$ and $\Delta \lambda$. To this end, we plot $\mathcal{S}(N,g)$ as a function of $(g-g_c)N^{1/\nu}$ for different system sizes $N$, where $\mathcal{S}$ denotes one of these quantities. In Fig.~\ref{fig:app2}(a)-(e) we show the data collapse for $\alpha = 2$ and ferro- and antiferromagnetic interactions. In order to test the quality of the collapse we further compute the $\chi^2$ value. For that we take the scaling function $\varphi_{\mathcal{S}} \equiv \phi_{\mathcal{S}}((g-g_c)N^{1/\nu})$ with $N=128$ obtained from the numerical data and linear interpolation as a reference. The deviation of the scaling functions is then measured by
\begin{equation}
\chi^2 = \sum_{i} \sum_{N} \frac{\left(\varphi_{\mathcal{S}}(x_{i,j})-\phi_{\mathcal{S}}(x_{i,j})\right)^2}{\phi_{\mathcal{S}}(x_{i,j})}
\end{equation}
where $x_{i,j} = (g_i - g_c)N_j^{1/\nu}$. Here the sum runs over various values $g_i$ and we choose $N = 32,46,64,90,128,256$ for ferromagnetic and $N = 16,22,32,46,64,90$ for antiferromagnetic couplings. In total, we consider around $250$ ($350$) distinct values $x_{i,j}$ in the range $[-10,10]$. The resulting values for $\chi^2$ are shown in Fig.~\ref{fig:app2}(f). The obtained quality of the collapse is very good, except, again, for the critical exponent $z$ for $\alpha = 0.2$ and antiferromagnetic interactions. Thus, larger system sizes than those considered here are required for such long-range interactions.

%\bibliographystyle{apsrev4-1}
%\bibliography{paper.bib}

\end{document}